\documentclass{article}

\usepackage[final, nonatbib]{nips_2017}

\usepackage[utf8]{inputenc} 
\usepackage[T1]{fontenc}    
\usepackage{hyperref}       
\usepackage{url}            
\usepackage{booktabs}       
\usepackage{amsfonts}       
\usepackage{nicefrac}       
\usepackage{microtype}      
\usepackage{algorithm}
\usepackage[noend]{algpseudocode}
\usepackage{graphicx}
\usepackage{subfig}
\usepackage{bm}
\usepackage{multirow}
\usepackage{dsfont}
\usepackage{wrapfig}

\title{Precision Scaling of Neural Networks for Efficient Audio Processing}

\author{
  Jong Hwan Ko \\
  School of Electrical and Computer Engineering\\
  Georgia Institue of Technology\\
  \texttt{jonghwan.ko@gatech.edu} \\
 \And
Josh Fromm \\
  Department of Electrical Engineering\\
  University of Washington\\
  \texttt{jwfromm@uw.edu} \\
 \And
  Matthai Philipose, Ivan Tashev, and Shuayb Zarar\\
  Microsoft AI and Research\\
  \texttt{\{matthaip, ivantash, shuayb\}@microsoft.com} \\
}

\begin{document}

\maketitle

\begin{abstract}
  While deep neural networks have shown powerful performance in many audio applications, their large computation and memory demand has been a challenge for real-time processing. In this paper, we study the impact of scaling the precision of neural networks on the performance of two common audio processing tasks, namely, voice-activity detection and single-channel speech enhancement. We determine the optimal pair of weight/neuron bit precision by exploring its impact on both the performance and processing time. Through experiments conducted with real user data, we demonstrate that deep neural networks that use lower bit precision significantly reduce the processing time (up to 30x). However, their performance impact is low (< 3.14\%) only in the case of classification tasks such as those present in voice activity detection.
\end{abstract}

\section{Introduction}

Voice activity detection (VAD) and speech enhancement are critical front-end components of audio processing systems, as they enable the rest of the system to process only the speech segments of input audio samples with improved quality \cite{Zhang2016}. With the rapid development of deep-learning technologies, VAD and speech enhancement approaches based on deep neural networks (DNNs) have shown powerful performance highly competitive to conventional methods \cite{Tashev2016, Xu2014, Zhang2013}. However, DNNs are inherently complex with high computation and memory demand \cite{7926982}, which is a critical challenge in real-time speech applications. For example, even a simple 3-layer DNN for speech enhancement requires 28 MOPs/frame and 56 MB of memory, as shown in column 3 of Table \ref{table:1}.

\begin{table}[h]
  \centering
  \caption{  Computation/memory demand and performance of DNNs with baseline/reduced bit width. The processing time was measured on a CNTK framework \cite{Yu2014} with an Intel CPU.}
  \centerline{\includegraphics[trim={0 8.5cm 0 6.3cm},clip, width=1\linewidth]{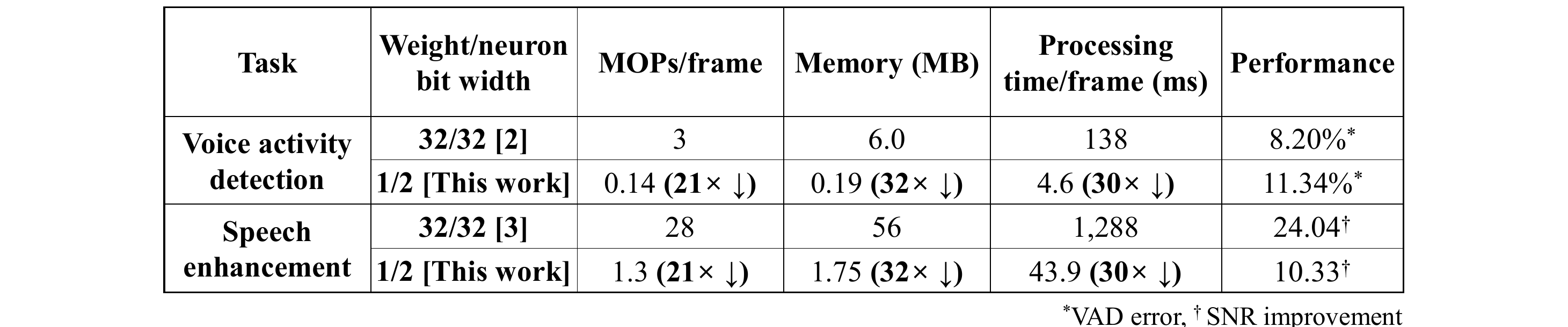}}

    \label{table:1}
\end{table}

A recently proposed method for reducing the computation and memory demand is a precision scaling technique that represents the weights and/or neurons of the network with reduced number of bits \cite{Hubara2000}. While several studies have shown effective application of binarized (1-bit) networks in image classification tasks \cite{Wu2016, Naetal.}, to the best of our knowledge, no work has been done to analyze the effect of various bit-width pairs of weights and neurons on the processing time and the performance of audio processing tasks like VAD and single-channel speech enhancement.

In this paper, we present the design of efficient deep neural networks for VAD and speech enhancement that scales the precision of data representation within the neural network. To minimize the bit-quantization error, we use a bit-allocation scheme based on the global distribution of the weight/neuron values. The optimal pair of weight/neuron bit precision is determined by exploring the impact of bit widths on both the performance and the processing time. Our best results show that the DNN for VAD with 1-bit weights and 2-bit neurons (W1/N2) reduces the processing time by $30\times$, providing $3.7\times$ lower processing time and 9.54\% lower error rate than a state-of-the-art WebRTC VAD \cite{webrtc}. For speech enhancement, the DNN with W1/N2 bit precision enhances SNR (signal-to-noise ratio) by 10.33 with $30\times$ smaller processing time.

\section{Precision Scaling of Deep Neural Networks}

\begin{figure}[b]
  \centering
  \subfloat[Example bit allocation]{\includegraphics[trim={0.0cm 0.0cm 0.0cm 0.0cm},clip, width=.50\textwidth]{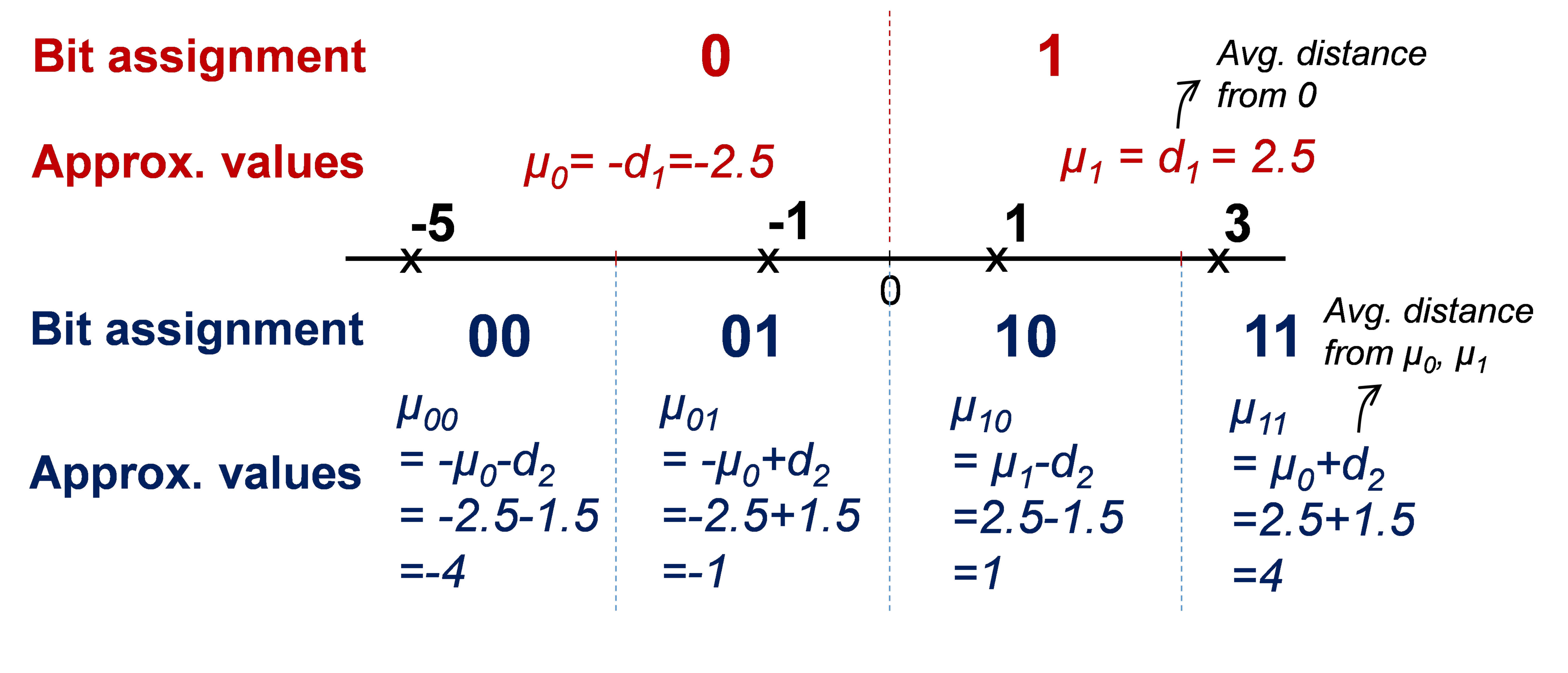}}\hfill
  \subfloat[({\bf Top}) 32-bit, ({\bf Bottom}) 1-bit network.]{\includegraphics[trim={0.0cm 0.0cm 0.0cm 0.0cm},clip, width=.38\textwidth]{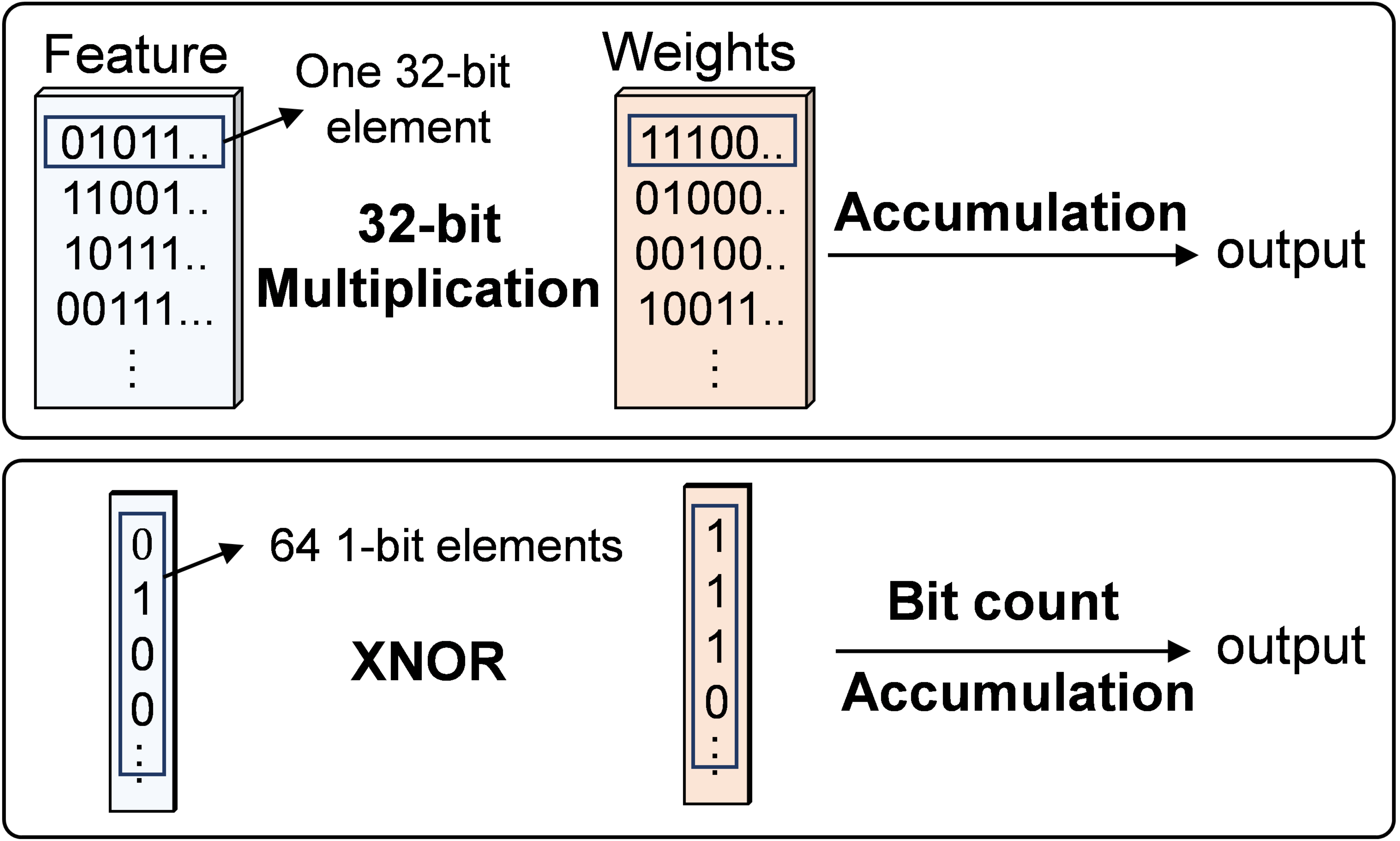}}\hfill
  \caption{The approach of extreme precision scaling or binarization of neural networks that is distribution sensitive.}
  \label{fig:control}
\end{figure}

While the rounding scheme is commonly used for precision scaling \cite{Gupta2015}, it can result in large quantization error as it does not consider global distribution of the values. In this work, we use a precision scaling method based on residual error mean binarization \cite{tang2017train}, in which each bit assignment is associated with the corresponding approximate value determined by the distribution of the original values. As illustrated in Figure \ref{fig:control}(a), the first representation bit is assigned deterministically based on their sign, and the approximate value for each bit assignment is computed by adding/subtracting the average distance from the reference value (0 in the first bit assignment). Each approximate value becomes the reference of each bit segment in the next bit assignment step. This approach allocates the same number of values in each bit assignment bin to minimize the quantization error.

\begin{figure}[t]
  \centering
  \centerline{\includegraphics[trim={0 0.2cm 0 0.2cm},clip, width=0.60\linewidth]{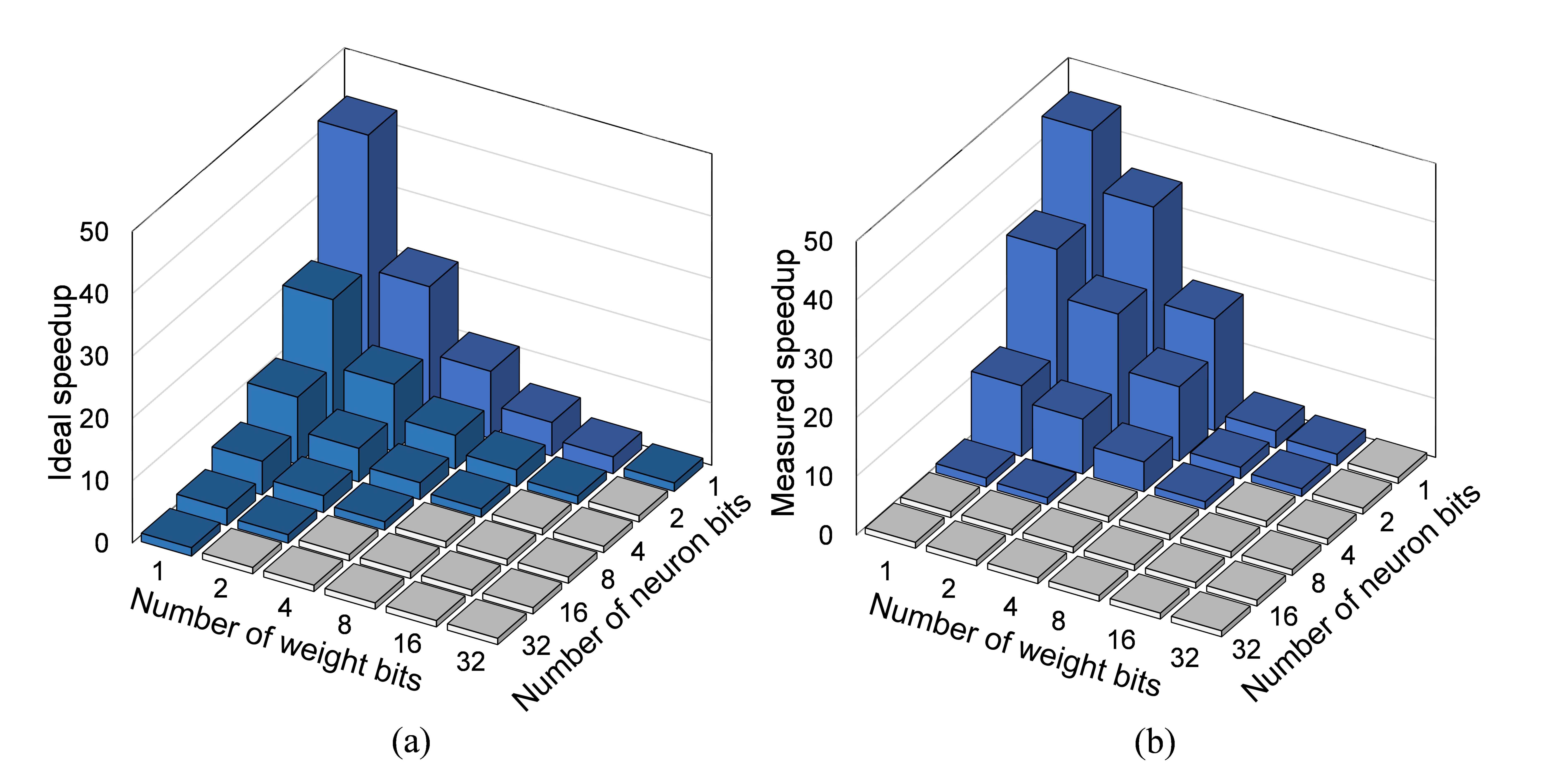}}
  \caption{ Speedup due to reduced bit precision of neurons and weights: (a) Ideal and (b) measured speedup. Blue bars indicate speedup > 1 and gray bars indicate speedup = 1.
}
    \label{fig:speedup}
\end{figure}

We estimate the ideal inference speedup due to the reduced bit precision by counting the number of operations in each bit-precision case [see Figure \ref{fig:control}(b)]. In the regular 32-bit network, we need two operations (32-bit multiplication and accumulation) per one pair of input feature and weight element. When the network has 1-bit neurons and weights, multiplication can be replaced with XNOR and bit count operations, which can be performed with 64 elements per cycle. When the network has 2 or more bit neurons and weights, we need to perform the three operations for all combinations of the bits. Therefore, the ideal speedup is computed as

$$ Speedup = max \Bigg(1,  \frac{128}{3 \times weight\ bit \ width \times neuron\ bit\ width } \Bigg).$$

We have implemented our precision scaling methodology within the CNTK framework \cite{Yu2014}, which provides optimized CPU-implementations for variable bit precision DNN layers. Figure \ref{fig:speedup} shows the ideal speedup and the actual speedup  measured on an Intel processor. The measured speedup is similar to or even higher than the ideal values because of the benefits of loading the low-precision weights, as the bottleneck of the CNTK matrix multiplication is memory access. The figure also indicates that reducing weight bits leads to higher speedup than reducing neuron bits since the weights can be pre-quantized, making their memory loads very efficient.

\section{Experimental Framework}

\begin{figure}[b]
  \centering

  \subfloat[VAD]{\includegraphics[trim={0.0cm 0.0cm 0.0cm 0.0cm},clip, width=.45\textwidth]{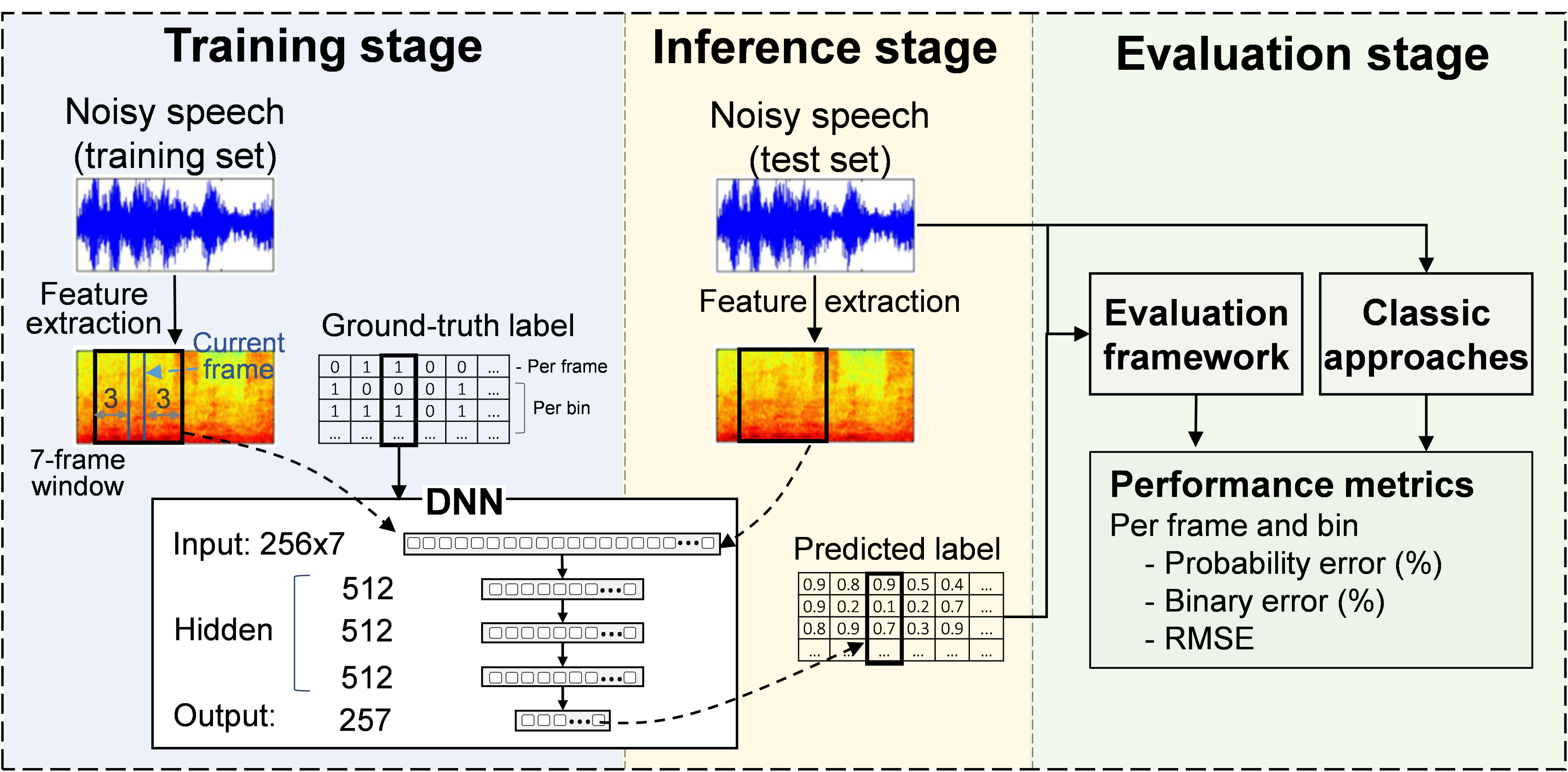}}\hfill
  \subfloat[Speech enhancement]{\includegraphics[trim={0.0cm 0.5cm 0.0cm 0.0cm},clip, width=.445\textwidth]{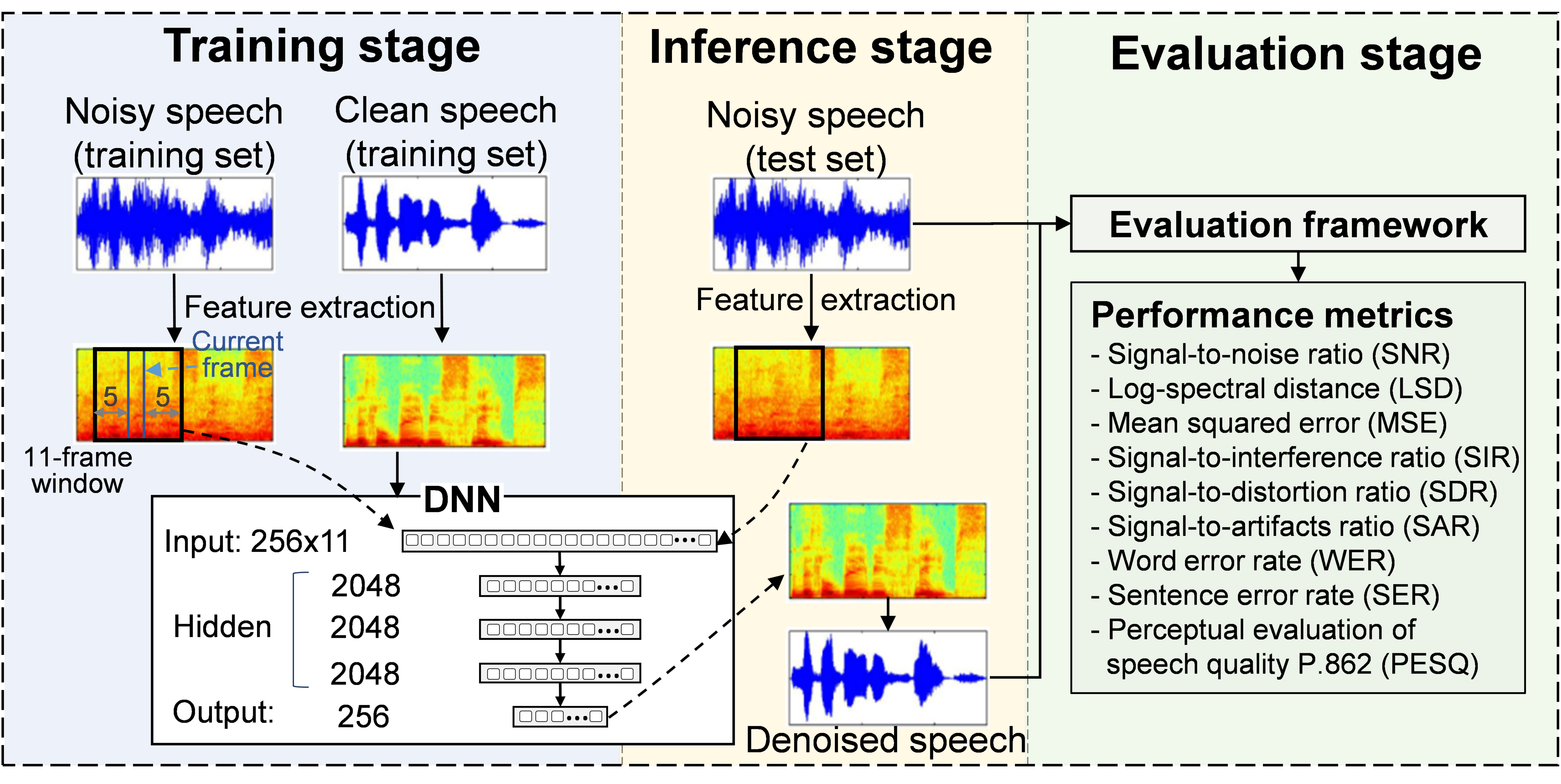}}\hfill

  \caption{Experimental framework.}
  \label{fig:framework}
\end{figure}

{\bf Dataset:} We created 750/150/150 files of training/validation/test datasets by convolving clean speech with room impulse responses and adding pre-recorded noise at different SNRs and distances from the microphone. Each clean speech file included 10 sample utterances that were collected from voice queries to the Microsoft Cortana Voice Assistant. Further, our noise files contained 25 types of recordings in the real world.

{\bf VAD:} As shown in Figure \ref{fig:framework}(a), we utilized noisy speech spectrogram windows of 16 ms and 50\% overlap with a Hann smoothing filter, along with the corresponding ground-truth labels for DNN training and inference. Our baseline DNN had three 512-neuron hidden layers with 7-frame windows as in \cite{Tashev2016}. The network was trained to minimize the squared error between the ground-truth and predicted labels. Then the noisy spectrogram from the test dataset was used to generate the predicted labels, which were compared with the ground-truth labels to compute performance metrics.

{\bf Speech enhancement:} The framework we used in this case was similar to the one for VAD, except for the use of clean speech spectrogram for training instead of the ground-truth activity label. We utilized the baseline DNN model with three hidden layers presented in \cite{Xu2014}. After performing the inference, the denoised speech from the output layer was used to compute the list of performance metrics shown in Figure \ref{fig:framework}(b). Due to space limitations, and since they are good proxies for speech quality, in this paper we only discuss the SNR and PESQ \cite{PESQ} metrics.

\section{Experimental Results}

\begin{figure}[t]
  \centering
  \centerline{\includegraphics[trim={0 5.8cm 0 6.8cm},clip, width=0.89\linewidth]{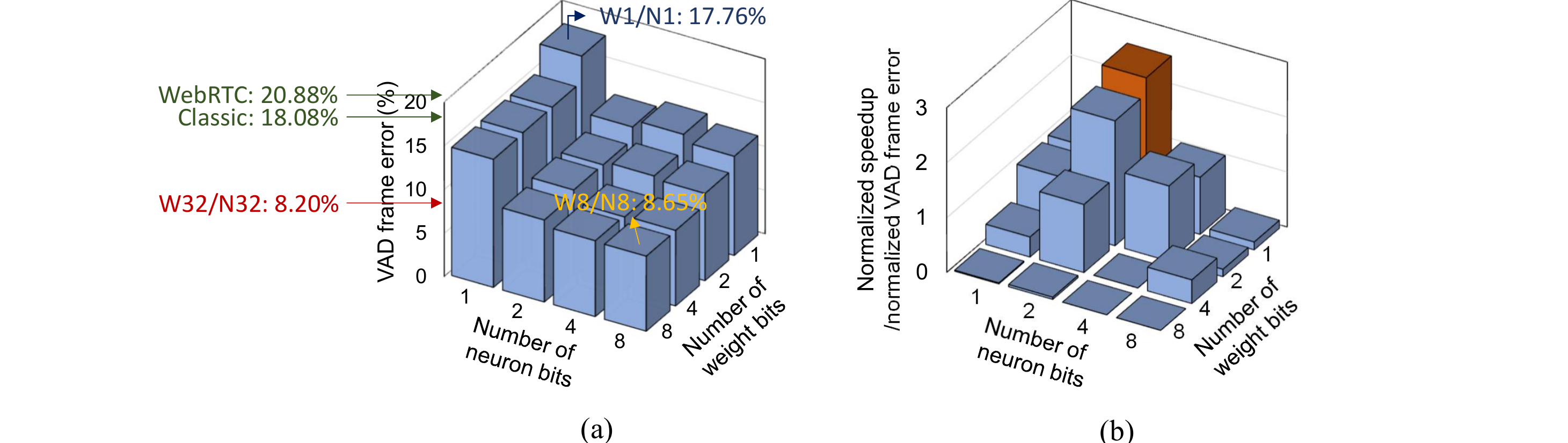}}
  \caption{ VAD performance of DNN with different pairs of weight/neuron bit precision. (a) Frame-level binary detection error and (b) normalized speedup/normalized VAD frame error.  A red bar indicates the optimal pair of bit precision (1-bit weights/2-bit neurons).
}
    \label{fig:VAD_result}
\end{figure}

\begin{figure}[b]
  \centering
  \centerline{\includegraphics[trim={1.5cm 6cm 3cm 6.8cm},clip, width=0.80\linewidth]{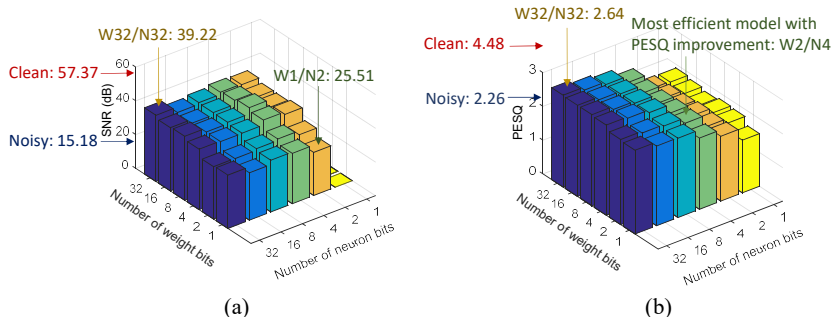}}
  \caption{ Speech enhancement performance of DNN with different precision. (a) SNR and (b) PESQ.}
    \label{fig:Speech_result}
\end{figure}

{\bf VAD:} Figure \ref{fig:VAD_result}(a) indicates that the detection accuracy of the DNN is more sensitive to neuron bit reduction than weight bit reduction. Note that even the DNN with 1-bit weights and neurons provides lower detection error than non-DNN based methods such as classic VAD \cite{Tashev2009} and WebRTC VAD \cite{webrtc}. To choose the optimal pair of weight/neuron bit precision in terms of detection accuracy and processing time, we introduce a new metric computed by multiplying normalized speedup and VAD error. Figure \ref{fig:VAD_result}(b) shows that the optimal bit precision pair is determined as 1-bit weights and 2-bit neurons (W1/N2). As we reduce the bit width to W1/N2, the per-sample processing time reduces from 138 ms to 4.6 ms ($30\times$ reduction), with a slight increase in the error rate (8.20\% to 11.34\%). The DNN with W1/N2 outperforms the WebRTC VAD with $3.7\times$ lower processing time and 9.54\% lower error rate.

{\bf Speech enhancement:} As Figure \ref{fig:Speech_result}(a) shows, SNR is improved for all bit-width pairs, except for 1-bit neurons. The optimal bit precision pair considering inference speedup and SNR improvement is W1/N2. However, Figure \ref{fig:Speech_result}(b) shows that the PESQ improvement is not achieved by DNNs with low bit precision; the most efficient model that enhances PESQ is W2/N4 with $9\times$ speedup. This is mainly because of the limited capability of the baseline DNN model, which improves PESQ by 0.38. The result also indicates that the lower-precision values (especially in the neural bit) are not suitable for an estimation or regression task (such as speech enhancement).

\section{Conclusions}
In this paper, we presented a methodology for efficiently scaling the precision of neural networks for two common audio processing tasks. Through a careful design-space exploration, we demonstrated that a DNN model with optimal bit-precision values reduces the processing time by $30\times$ with only a slight increase in the error rate. Even at these modest precision scaling levels, it outperforms a state-of-the-art WebRTC VAD with $3.7\times$ lower processing time and 9.54\% lower error rate. The low bit precision DNN also enhances the quality of noisy speech, but the precision could not be reduced much for speech enhancement. Our results indicate that the precision scaling of DNNs may be better suited for classification or detection tasks such as VAD rather than estimation or regression tasks such as speech enhancement. To validate this hypothesis, we intend to further explore the scaling of neural-network bit precisions for other classification tasks such as source separation and microphone beam forming and estimation tasks such as acoustic echo cancellation.


\small

\bibliographystyle{unsrt}   

\end{document}